# EFFECT OF PRE-OXIDATION TREATMENTS ON THE STRUCTURAL, MICROSTRUCTURAL, AND CHEMICAL PROPERTIES OF (Ni,Pt)Al SYSTEMS

J.E. Garcia-Herrera[1], D.G. Espinosa-Arbeláez[2], L.A. Caceres-Díaz[1], G.C. Mondragón-Rodríguez[2], H. Ruiz-Luna[3], J. Gonzalez-Hernandez[2], J. Muñoz-Saldaña[4] and J.M. Alvarado-Orozco[2,5*]

[1]CONACYT - Centro de Tecnología Avanzada, CIATEQ, Eje 126 No.225, Industrial San Luis, San Luis, 78395, México.
[2]Centro de Ingeniería y Desarrollo Indutrial (CIDESI), Av. Playa Pie de la Cuesta No. 702, Desarrollo San Pablo, Querétaro, 76125, México.
[3]CONACYT – Universidad Autónoma de Zacatecas, 98000, Zacatecas, México
[4]Centro de Investigación y de Estudios Avanzados del IPN, Unidad Querétaro, Libramiento Norponiente 2000, Real de Juriquilla, 76230, Querétaro, México.
[5]Consorcio de Manufactura Aditiva, CONMAD, Av. Pie de la Cuesta 702, Desarrollo San Pablo, Querétaro, México.

## Abstract

The effect of isothermal pre-oxidation treatments on the β-(Ni,Pt)Al + René N5 system degradation is here reported. The oxidation treatments were carried out from 900 (mostly θ-$Al_2O_3$ growing conditions) to 1200°C (mainly α-$Al_2O_3$ growing conditions) for 5 h, under a purified Ar-stream with an fixed $pO_2$= 1 x $10^{-5}$ atm. Results are discussed based on the correlation between the structural, microstructural and chemical properties of the β-(Ni,Pt)Al BC showing that pre-oxidation parameters have an important effect on the multi-elemental counter diffusion phenomena along BC. For instance, a significant BC+IDZ thickness increase of 55% at 1200 °C was observed with respect to as-received sample just after 5 h of oxidation resulting in a severe BC degradation.

***Corresponding author**: Tel. +52 442 2119800 ext. 5037, Fax. +52 442 2119938

email: juan.alvarado@cidesi.edu.mx

1. **Introduction**

Thermal Barrier Coating (*TBC*) systems have been used for decades to protect hot-section components of gas turbines [1,2]. *TBC* systems are multi-layered coatings deposited on a superalloys (*SA*) substrate and contain: a) an alumina-forming metallic alloy bond coat (*BC*), *e.g.*, NiPtAl or NiCoCrAlY, b) a thermal growth oxide (*TGO*) as result of *BC* oxidation, and c) a ceramic top coat (*TC*) as a thermal insulator, typically 7 wt.% $Y_2O_3$ stabilized $ZrO_2$, known as 7YSZ. The lifetime of *TBC* systems are mainly limited by intrinsic mechanisms associated with the interplay between *BC*/*TGO*/*TC* and their interfaces [3]. Failure of *TBC* systems is usually manifested as *TC* spallation. Because of this, efforts have been made to improve the lifetime of *TBC* systems using various strategies, including pre-oxidation treatments prior 7YSZ-*TC* deposition in order to improve the *TGO* properties [4–7].

Pre-oxidation can be defined as a heat treatment of the *BC* + *SA* system controlling the oxygen partial pressure ($pO_2$), temperature and time to promote and control the growth of a continuous and stable aluminum oxide scale (*i.e.* $\theta$-$Al_2O_3$, $\alpha$-$Al_2O_3$) prior *TC* deposition [8–10]. In previous experiments done in our group, the polymorphic evolution of the $Al_2O_3$ upon temperature was analyzed. The formation of $\theta$-$Al_2O_3$ phase takes place at 900°C and starts to transform into the $\alpha$-$Al_2O_3$ phase at slightly higher temperature (*ca.* 950°C), reaching a full transformation into $\alpha$-$Al_2O_3$ after ~ 4 h oxidation at 1150°C in Argon (*i.e.*, $pO_2$= 1x $10^{-5}$ atm) atmosphere [8]. A separate study showed that low-$pO_2$ sluggish the $\theta$-$Al_2O_3$ growth (*p*-type oxide) according with the relation $k_{p\theta} \propto p_{O_2}^{3/16}$ [10].

Furthermore, it has been reported that a good quality $\alpha$-$Al_2O_3$ scale with controlled grain size and homogeneity will lead to an improvement adhesion of the *BC*/*TGO*/*TC* interfaces [5,9,10]. However, there is a lack of information about how pre-oxidation can affect the physical and chemical properties of the *BC*+*SA* system because of the elemental diffusion along the layers during pre-oxidation treatments. This information is essential to select the proper pre-oxidation parameters to avoid premature or induced a *BC* degradation during this treatment. For instance, Al is thermally activated during pre-oxidation and diffuses either, to the surface (outward diffusion), leading to the formation of an alumina scale, or inward-diffusion to the *SA*. Additionally, due to the existing chemical gradients of several elements between the β-(Ni,Pt)Al *BC* and the René N5 *SA*, diffusion of Pt takes place from the *BC* to the *SA* (inward-diffusion) and heavy elements (*e.g.*, Co, Cr, Ta, W, Mo) diffuse from the *SA* to the *BC* [11]. Based on this, the objective of this work is to investigate the effect of isothermal pre-oxidation treatments on the physical and chemical properties of the β-(Ni,Pt)Al+René N5 system. Specifically, the correlation with the structural, microstructural, and chemical properties after thermal exposures.

**Material and methods**

Rectangular samples (1.8 x 1.2 x 0.15 cm) were provided by GE Aircraft Engines (Evendale, OH). The β-(Ni,Pt)Al *BC* was deposited by the Johnson-Matthey method on a grit-blasted single-crystal René N5 *SA*. This geometry was required for the *in-situ* oxidation kinetics study, which has been reported elsewhere [8,10]. In the present contribution, oxidations exposures were carried out to investigate the β-(Ni,Pt)Al+René N5 system degradation in the range of 900 to 1200°C (with 50°C increments) for 5 hours, under purified argon stream (working gas) with a $pO_2 = 1 \times 10^{-5}$ atm, and with a total working pressure of 1 atm. To control the desired oxygen concentration, the argon was passed through a gas purification system (OG-120M, Oxygon Industries, USA). After pre-oxidation treatments, the samples were subsequently cross-sectioned and carefully polished. Preparation started with abrasive wet cutting using a low-speed, diamond-wafer blade on an automatic cut-off machine. Samples were mounted in pairs, using a low-viscosity cold-setting epoxy, with the oxidized surfaces facing each other to avoid edge rounding. The mounted samples were grounded using a series of water-cooled silicon carbide abrasive papers down to a 4000-grit finish (*i.e.*, according with the Federation of the European Producers of Abrasives which is equivalent to the 1200 standard ANSI grit).

To reveal the microstructure, the samples were mechano-chemically polished with a commercial colloidal silica suspension (MasterMet2, Buehler, USA) followed by chemical etching using a 5 vol.% HF aqueous solution. Finally, the samples were ultrasonically cleaned in methanol and deionized water for 20 min and dried with nitrogen gas.

The chemical analysis was conducted in the cross section of the samples using a FEG-SEM (Quanta 3D 200i FEI, NL) equipped with an EDX-spectrometer (INCA 350 EDS, Oxford Instruments, UK) and EPMA (JXA-8900R WD/ED Microanalyzer, JEOL, Japan) with the purpose of correlating the local chemical composition with the phase transformations of the β-(Ni,Pt)Al *BC* as a function of the thermal treatments. These were also followed by X-ray diffraction analysis using a Rigaku Dmax/2100 vertical diffractometer with a Cu-kα radiation, XRD details.

2. **Results and discussion**

The microstructure of the β-(Ni,Pt)Al + René N5 system pre-oxidized at 1050 and 1200ºC for 5 h is shown in Figure 1. An image of the cross section of the sample in the as-received condition is also shown as a reference to compare with the effects promoted by the heat treatments at the intermediate and highest temperature during the pre-oxidation experiments. Owing to the interdiffusion processes at the *TGO/BC* and *BC/SA* interfaces, a clear increase of the *BC+IDZ* thickness is observed with temperature. This increase is about 15 % for the sample treated at 1050 ºC (Fig. 1b), and about 55% for the sample treated at 1200ºC (Fig. 1c) with respect to the thickness of the as-received sample (Fig. 1a). Furthermore, after 5 h of exposure at 1200°C, the γ´-phase is observed at the *BC* grain boundaries (Fig. 1c) as an evidence of the *BC* Al-depletion. It is also noticed, the coarsening and growing of the *IDZ* precipitates as a result of the *BC+IDZ* growth and the low solubility of refractory metals in the β phase [12,13].

The presence of γ´ phase was confirmed by XRD analysis and it is discussed below. No significant changes were observed in the *BC+IDZ* thickness for samples oxidized below 1050 ºC.

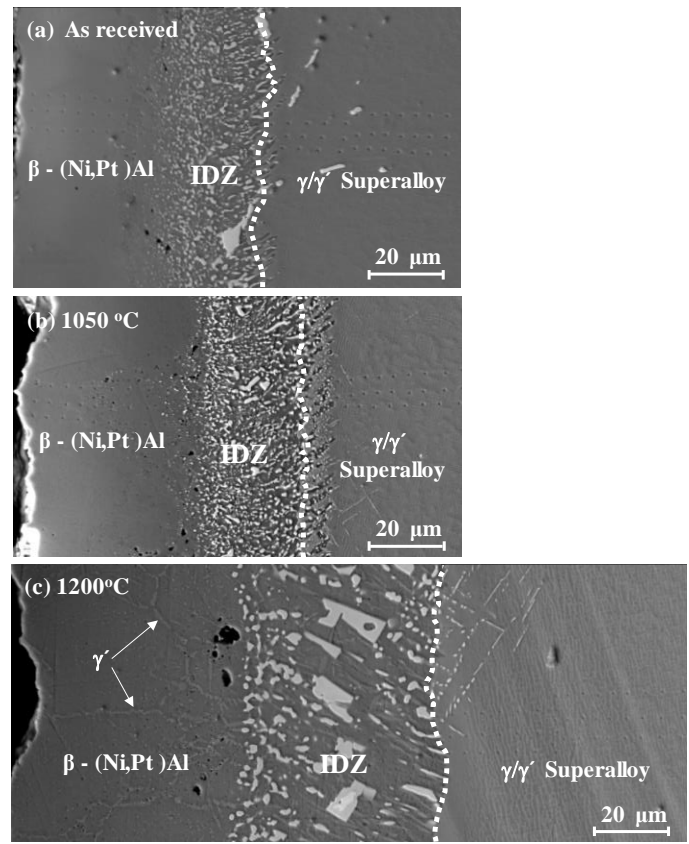

*Figure 1. Cross section SEM images of the (a) as-received sample (substrate + bond coat system), and two representative pre-oxidation conditions after the treatments at (b) 1050 ºC and (c) 1200 ºC for 5h in $pO_2=1 \times 10^{-5}$ atm.*

Changes in chemical composition along the cross section of the β-(Ni,Pt)Al +IDZ+ René N5 system are shown in the Figure 2 for the samples oxidized at 900 and 1200ºC, and compared with the sample in the as-received condition. These temperatures were selected as known conditions where the *TGO* grow kinetics is mainly controlled by the θ-$Al_2O_3$ and α-$Al_2O_3$ phases, respectively [8, 10].

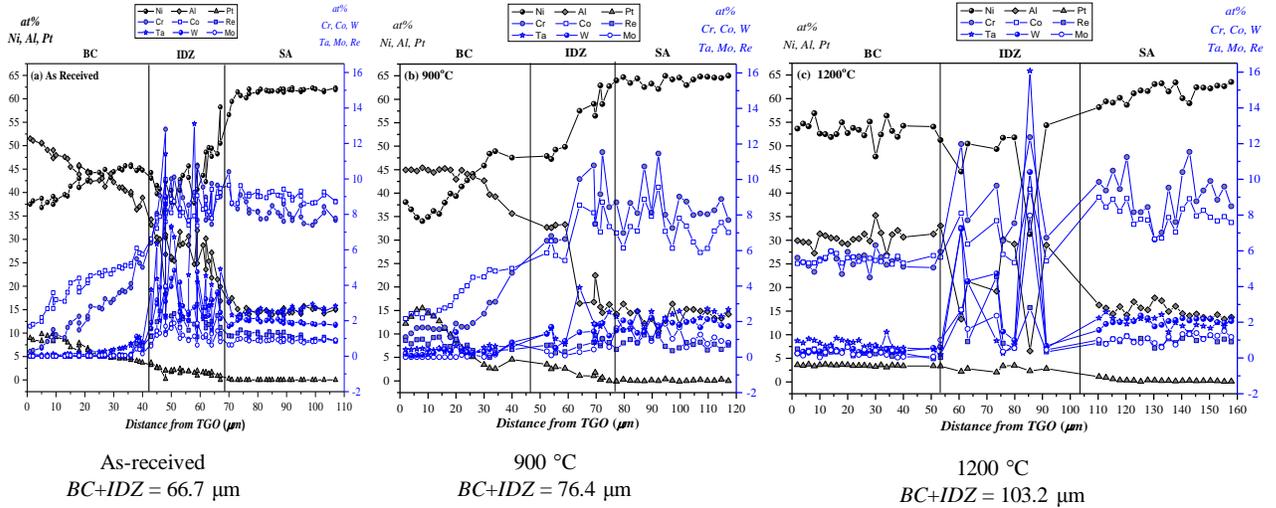

| As-received | 900 °C | 1200 °C |
| --- | --- | --- |
| BC+IDZ = 66.7 μm | BC+IDZ = 76.4 μm | BC+IDZ = 103.2 μm |

*Figure 2. Chemical composition profiles along the system: bond coat, interdiffusion zone and superalloy in (a) as-received condition, and after pre-oxidation treatment for 5h in $pO_2 = 1 \times 10^{-5}$ atm at (b) 900°C and (c) 1200°C.*

In the as-received sample (Fig. 2a), it is observed that the elements Al, Ni, Pt, Cr and Co varied almost linearly along the first 40 μm in the *BC* owing to the manufacturing process. The Al and Pt content gradually decreased from 51.5 to 38.9 at. % and 8.9 to 3.7 at. %, respectively. On the contrary, the Ni, Cr and Co percentages increase from 37.5 to 45.1, 0.2 to 5.0, and 1.7 to 5.8 at. %, respectively. The *IDZ* shows heterogeneous concentrations of each element and in the *SA* region no changes in the chemical composition were observed. For the as-received and tested conditions, the percentage values of the refractory elements such as Re, Ta, W y Mo remain close to 0 at% (*i.e.*, right-hand side scale).

According to the analysis, chemical composition gradients are observed from the *BC* surface to the *IDZ* boundary. These gradients are result of the coating deposition process, which starts with a Pt-electroplating on a grit-blasted SA substrate, followed by a heat treatment step to promote the Pt diffusion [8]. Later, a low-activity aluminum coating is deposited by chemical-vapor deposition (CVD) technique to finish the *BC* [14].

During the low-activity process, Ni diffuses outward from the *SA* and reacts with the Al vapor species present in the chamber atmosphere resulting in the formation of the β-(Ni,Pt)Al *BC* [15]. This process also allows the outward diffusion of small amounts of Co and Cr from the *SA*.

Figure 2b shows the changes in β-(Ni,Pt)Al *BC* chemical composition after 5 h oxidation at 900ºC. Along the first 25 µm, the gradient composition of Al is about zero, with an Al-content of approximately 45 at. %. Moreover, a continuous Pt-enrichment is observed as a result of the Ni inward diffusion caused by the TGO formation, reaching a maximum Pt content of about 16 at. % at about 8 µm.

For the sample pre-oxidized at 1200°C (Fig. 2c), clear diferences within the *BC* zone are observed in comparison with the as-received and 900 °C conditions. The concentration of Al and Pt decrease, while the Ni, Cr and Co increase and reach a constant chemical composition along the *BC*. This specimen shows the lowest Al and Pt contents and the highest for Ni, while the gradients of Cr and Co show minimal variations.

According to the microstructural and chemical composition results, the chemical gradients are clearly and highly dependent on the temperature of the pre-oxidation treatment. As the temperature increases, greater are the changes occurring in the system (Figs. 1 and 2). Based on the chemical analysis, a higher Al-depletion percentage was observed for the *BC* pre-oxidized at 1200ºC than for the sample oxidized at 900ºC. Previous studies concerning the oxidation of β-(Ni,Pt)Al *BC* systems have found that at 1200ºC is α-$Al_2O_3$, the phase controlling the *TGO* oxidation kinetics [10]; It is important to mention that an accelerated Al-depletion at high temperatures is undesirable because this can lead to a premature failure or degradation of the system.

Given that diffusion is a thermally activated process, during the pre-oxidation treatments two diffusion mechanisms occur. Aluminum within the β-(Ni,Pt)Al phase outward diffuses from the

*BC* surface to form the *TGO*, while Al and Pt inward diffuse to the superalloy as a result of the chemical potential gradients that promote the BC and IDZ growth.

The inward diffusion is strengthened due to the *TGO* formation which acts as a diffusion barrier against the elemental species in the system, including the Al. It is important to notice that, even when pre-oxidation treatments promote *TGO* growth, if the exposure treatments are not suitable conditions, they will also generate a *BC* degradation as a consequence of the loss Al and Pt content which reduces the *BC* lifetime (see Fig. 2b-c), this is agreement with the reported by Chen et al. [13].

The *BC* degradation is mainly affected by counter-diffusion between *BC* and *SA* [16-17]. An evidence of this is clearly observed in Fig. 2b. The stabilization of the Al content during the first 25 μm of the *BC* shows that $Al_2O_3$ scale acts as a diffusion barrier at the *TGO*/*BC* interface and the next 15 μm the elemental gradient in *BC* shows the inward diffusion of Al until the *IDZ*/*SA*. This is, the Al depletion in the *BC* is mainly promoted by interdiffusion, and not by TGO growth.

A systematic study to investigate the effect of the oxidation temperature on the elemental composition within the *BC* is depicted in Fig. 3. A significant variation in the compositional gradients for the samples oxidized from 900 to 1050 °C were observed while at 1100 and 1200 °C a compositional steady-state is reached after 5 h. For the sample treated at 900°C, is notable a significant compositional change at the first 16 μm which is owing to the decrease of the Al and Ni content and its consequent Pt enrichment in comparison with as-received condition (Fig. 3a-c). The $Al_2O_3$ scale formation causes the Ni inward diffusion to compensate Al depletion which promotes an unbalance in the diffusivity of both elements, forming voids at metal-oxide interface [8,18,19], as can be clearly observed in Figure 4.

In this regard, it is well known that Pt retards the formation of mature voids injected due to the accumulation of Kirkendall vacancies and it occupies preferentially Ni sites in β-(Ni,Pt)Al alloy in comparison with non-Pt-containing alloys, *e.g.*, β-NiAl [18,20,21].

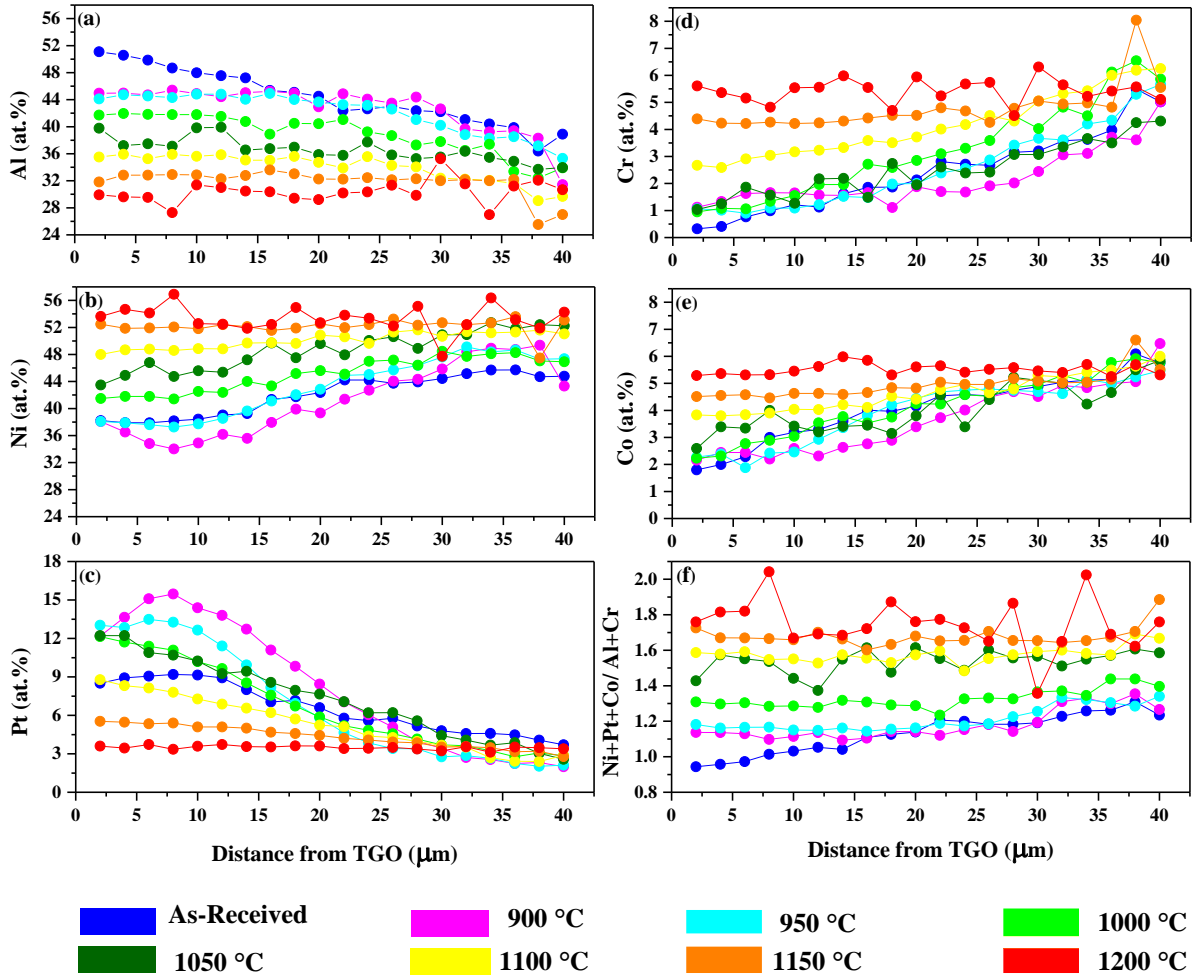

*Figure 3. Variation in the chemical composition of the bond coat along the cross section at the different temperatures tested.*

The decrease in Al content is also related with increased mobility caused by the presence of Pt which decreases the Al activation energy promoting the faster formation of an $Al_2O_3$ scale on the BC surface [20,22].

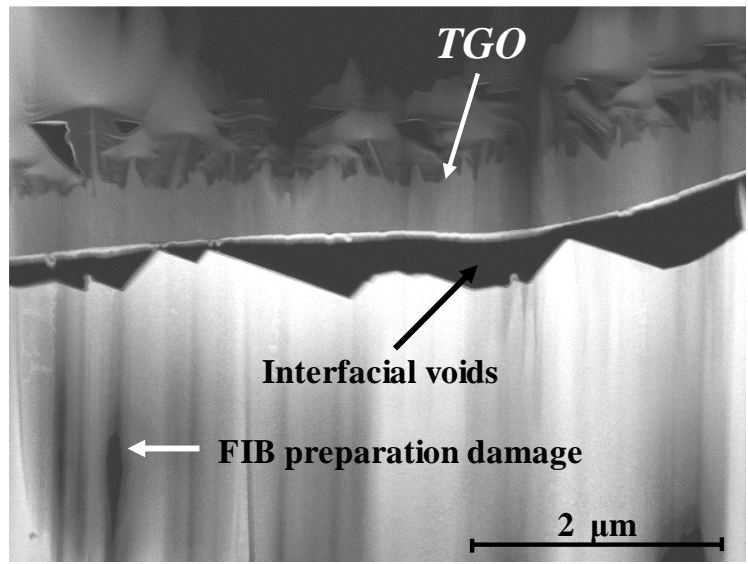

*Figure 4. Voids observed at the BC/TGO interface after pre-oxidation treatment at 900°C for 5 hours.*

Figs. 3b and 3d-e show how the Ni, Cr, and Co diffuse outward from the *SA* and hence, the gradient is positive regarding the distance. Since diffusion is a thermally activated process, the Ni, Cr and Co content are higher in the *BC* as the exposure temperature increases. For instance, at 900 °C the Ni, Cr and Co reaches 38.1, 1.0 and 2.1 at. % while at 1200 °C attains 53.7, 5.6 and 5.3 at. % respectively, at the *TGO/BC* interface.

The *BC* shows some interesting relationships. Ni and Al behave inversely as the pre-oxidation temperature increases. Also, it can be observed that the Ni changes in an inverse relationship with Pt, in a similar way than the Al behave with Cr. This behavior is explained in terms of the two-sublattice model of the ordered β−NiAl proposed by Ansara *et. al.*, in which the β phase is described by the δ and λ sublattices that correspond to the (0,0,0) and (½, ½, ½) wyckoff positions, respectively [23]. In the stoichiometric phase, Ni occupies the δ sublattice and Al occupies the λ sublattice positions. Deviations from stoichiometry are addressed by triple point defects to maintain the β crystal structure. Vacancies in the Al-rich side are created and Ni anti-sites appear in the Ni-rich side [24,25]. The elemental behavior in the coating can be understood considering the site preferences proposed by Jiang *et. al.*, Pt and Co have a strong preference for Ni sites in the β phase while the Cr site preference is dependent of the composition and temperature [26–28]. In the case of pre-oxidation experiments at high temperatures with subsequent Ni enrichment of the coating, Cr has a defined preference for Al sites in the λ sublattice [29]. The Fig. 3f shows the composition ratio of δ and λ sublattices considering the Pt, Co and Cr site preferences, respectively. It can be seen how the composition of the pseudo-binary (Ni+Pt+Co)(Al+Cr) moves from 1 to 1.9 approx., which is related to the change from the 50 to the 66 Ni fraction "$X_{Ni}=\delta/(\delta+\lambda)$" during the pre-oxidation experiments.

Based on this and considering a two-sublattice model, is possible to comprehend how the coating chemical composition shifts from stoichiometry to the boundary with β + γ' (L1$_2$) mixture region as a function of the thermal exposures. Figure 5 shows the BC chemical composition along thickness considering that Co atoms occupy the Ni sublattice whereas Cr atoms occupy the Al sublattice plotted in the isothermal section of the ternary Ni-Al-Pt phase diagram at 1100/1150 °C reported by Gleeson *et. al.* [30]. From right to left, the data represents the composition in the vicinity of the *TGO/BC* interface and along the *BC* until it reaches the *BC/IDZ* interface.

Despite this diagram is plotted at a specific temperature, it aims to show the correlation between the chemical composition and the phase transformations of the *BC* as a function of the temperature. Plotting the chemical compositions in this diagram at the different pre-oxidation temperatures, is an approach that allow observing the diffusion path and comparing to the structural evolution in the specimens. The as-received specimen composition (blue curve in Fig. 5) and those subjected to a heat treatment from 900 to 1000°C are located specifically in the β-NiAl (cubic structure) zone, irrespective of the variations in the chemical composition. As observed in the ternary diagram, the composition values of the sample treated at 1050°C are localized at the boundaries where the transformation from bbc β-phase to the tetragonal martensite L1$_0$-phase is expected to occur.

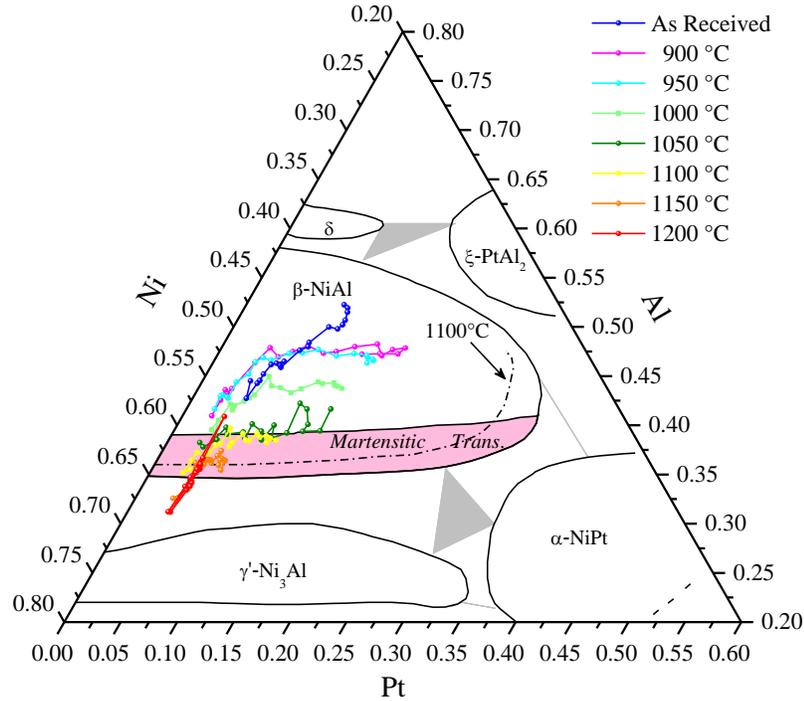

Figure 5. Ni-Pt-Al phase diagram at 1100/1150ºC. Normalized chemical composition variation of the BC along the cross section pre-oxidized at different temperatures. Reprinted from [30].

The β-phase and the martensitic region diminishes as the temperature also decreases as indicated by the dotted line in the phase diagram for a temperature of 1100ºC. The compositions values of the sample treated at 1100°C are located inside the martensitic transformation region while the composition for the specimens treated at 1150 °C and 1200 °C are placed in β + γ' phases mixture zone, without reaching the γ' phase in the ternary system. This result agrees with that observed in Fig. 1, where the γ' phase was identified in the 1200 ºC heat-treated specimen. Moreover, $X_{Ni}$ for the pseudo-binary (Ni+Pt+Co)(Al+Cr) system obtained by mean of sublattice composition normalization can be correlated to the observed diffusion path in Fig. 5. $X_{Ni}$ =50 corresponds to the ordered β phase of the as-received sample, while $X_{Ni}$ = 66 represents the Ni-enriched region where γ' is suitable to precipitate.

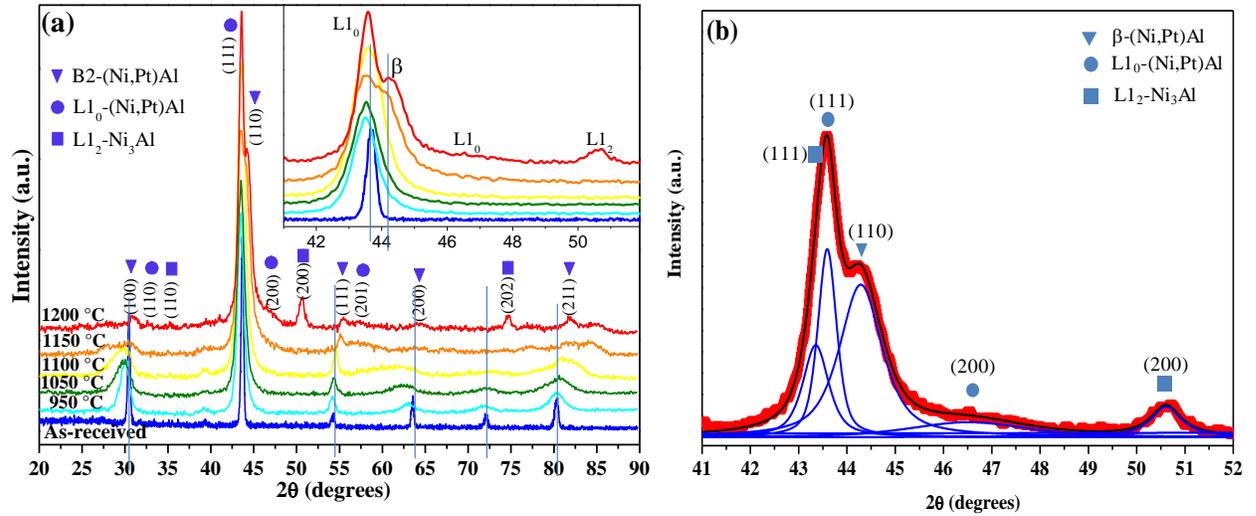

*Figure 6. (a) Structural evolution of the Bond Coats as a function of temperature treatment function from as-received to 1200 °C condition and (b) deconvoluted XRD pattern of the 1200 °C heat-treated sample.*

The XRD patterns of the specimens as a function of the pre-oxidation experiments are presented in Fig. 6a. The effect of the high temperature exposure on the crystal structure of the β-(Ni,Pt)Al *BC* is shown from bottom (as-received) to top (1200 °C). In the as-received conditions, the *BC* exhibits the characteristic superlattice (100), (111), (210) and fundamental (110), (200), (211) peaks. The fundamental peaks demonstrate the body centered cubic character of the *BC* and the superlattice peaks its ordered nature. As the exposure temperature increases up to 1100 °C, the β(100) and β(110) peaks shift to lower angles and the intensity of the superlattice (100) peak decreases. The shifting of the peaks corresponds to the increase in lattice constant of the β phase and may be associated to the change of Pt from 9 to 15 % at during the first 10 µm in the *BC*. This assumption considers that Pt has a remarkable site preference for the δ sublattice [28] and its atomic size is higher than the Ni atomic size (1.39 Å compared to 1.24 Å), then Pt replaces the Ni and contributes to the size of the β phase. The decrease of the intensity of the superlattice peaks shows the partial disordering of the β phase in the range of 950 °C to 1100 °C, and complete disorder from 1100 °C forward. It is observed that the β phase is disordered and transformed to the martensite phase after 1100 °C. The martensite phase is identified by the presence of the superlattice (110) and fundamental (111) and (200) peaks. This diffusion-less transformation is related to the Al depletion in the coating and it has been reported before for binary NiAl [16] and commercial β-(Ni,Pt)Al *BC* [12,31]. Smialek *et al.* [16] reported that the cubic β phase destabilizes and transforms to the tetragonal $L1_0$ phase for values below 37 at. % of Al. The onset of the martensitic transformation found in this work for 5 h exposure is between 1100 and 1150 °C with values of Al in the range of 33 to 36 at. %. These results are consistent with previously published data [31].

The XRD patterns deconvolution of the sample exposed to 1200 °C shown in the Figure 6b revealed the presence of the disordered β, $L1_0$ and gamma prime $L1_2$ phases. The $L1_2$ phase is recognized by the appearance of the superlattice (110) and fundamental (200), (202) peaks and it is the result of the diffusional transformation as the Al depletes. The XRD patterns allowed also to corroborate the diffusion path observed in the Fig. 5 when the normalization of the composition to the ternary Ni-Al-Pt system is considered. As can be seen in Figs. 5 and 6, the β phase is stable in the as-received (ordered β) and in the heat-treated sample exposed up to 1100 °C (partial disordered β). A mixture region of $L1_0$ + β can be found from 1100 °C to 1150 °C and a two-phase zone of $L1_2$ + disordered β is reached at temperatures above 1150 °C. These results are consistent with the microstructural observations and chemical compositions presented in Fig. 1c and Fig. 3, respectively. The phase transformations β → β + $L1_0$ and β → β + $L1_2$ that take place during the thermal exposures which imply that the pre-oxidation experiments carried out for 5 h caused a significant degradation of the *BC*. At this point, a critical balance among time, temperature, and oxide scales formation (θ- and/or α-$Al_2O_3$) must be considered during the pre-oxidation treatments of the *TBC* systems.

**Conclusions**

Pre-oxidation treatments were carried out in the β- (Ni, Pt) Al +IDZ+ René N5 system in order to investigate the effect of temperature on the structural, microstructural and chemical stability of the system. Multi-elemental counter diffusion phenomena were observed due to chemical potential along the β- (Ni, Pt) Al +IDZ+ René N5 system, where the most significant findings are described below:

- A significant BC+IDZ thickness increase from 15% to 55% in the range of 1050 °C to 1200 °C were observed with respect to as-received sample, as a result mainly of the elemental counter diffusion between BC and SA.
- The pre-oxidation treatments at temperatures below of 1050°C shown an unbalance diffusivity due to TGO-$Al_2O_3$ scale formation which promotes the Al depletion of the BC and the Ni inward diffusion causing a Pt-enrichment in the BC. For instance, a 16 at. % at about 8 µm for the sample pre-oxidized at 900°C with excessive interfacial voids formation at the *TGO*/*BC* interface as a result of the Kirkendall effect.
- The elemental flux is initially accelerated by the effect of temperature and subsequently attenuated by the growth of the *TGO*, which forms a diffusion barrier at the interface *BC*/*TGO*.
- The two-sublattice model adequately describes the elemental distribution within the crystalline phases and its phase evolution considering the study system. For instance, the *BC* phase transformation from β to β+γ' after 1100°C according with XRD.

Based on this findings, a final conclusion can be made. Despite pre-oxidation treatments have been proved to be beneficial for increasing the TBC systems lifetime due to the exclusive a-Al2O3 formation []; the pre-oxidation parameters (e.g., pO2, temperature and time among others) must be selected in such a way, that the pre-oxidation conditions promote the a-phase growth without compromising the BC integrity that may reduce the TBC system service performance.

**Acknowledgments**

The authors thank D. Konitzer at GE-Aviation, USA and Jose Luis Ortiz-Merino at General Electric Infrastructure Queretaro, Mexico for providing the samples used in this work. This project was supported by CONACYT, through the program *National Problems* and the project 2015-01-863, as well as the CATEDRAS Conacyt. The authors also thank the National Lab–CENAPROT and the consortiums CONMAD and MTH for providing all the facilities required for this work.